\documentstyle[aps,prl,floats,graphicx]{revtex}
\begin{document}
\draft
\twocolumn[\hsize\textwidth\columnwidth\hsize\csname @twocolumnfalse\endcsname
\title{Models for Superfluid $^3$He in Aerogel}
\author{E.V. Thuneberg$^{(1)}$, S.K. Yip$^{(2,3)}$, M. Fogelstr\"om$^{(2,3)}$ and
J.A. Sauls$^{(2)}$} 
\address{$^{(1)}$Low Temperature Laboratory, Helsinki University of 
Technology, Otakaari 3A, 02150 Espoo, Finland\\
$^{(2)}$Department of Physics and Astronomy, 
Northwestern University, Evanston, Illinois 60208
\\
$^{(3)}$Department of Physics, \AA bo Akademi, Porthansgatan 3, 20500 {\AA}bo, Finland}
\date{\today}
\maketitle
\begin{abstract}
Several recent experiments find 
evidence of superfluidity of $^3$He in 98\%-porous aerogel. The primary effect
of the aerogel is that it scatters the quasiparticles of $^3$He. We find that
many experimental findings are quantitatively understood by a relatively simple
model that takes into account strong inhomogeneity of the scattering on a length
scale of 100 nm. 
\end{abstract}
\pacs{PACS:  74.25.Fy, 74.70.Tx, 72.15.Eb}
\bigskip
] 

The discovery of unconventional paring states in high-temperature
superconductors has generated a lot of interest in impurity scattering in these
materials. In particular, the inhomogeneity of the scattering has been
considered recently \cite{FKBS}. However, both the experimental and theoretical
studies are  difficult because of the complicated structure of these
substances. Recently, a new possibility was opened for studying impurity
effects on unconventional pairing states: superfluid $^3$He in very porous
aerogel. This system has many advantages. For example, the pure state of
superfluid $^3$He is absolutely pure in experiments, and it is theoretically
very well understood.  A crucial parameter, the coherence length
$\xi_0$, can easily be varied within a factor 5 by varying the pressure.  The
torsional oscillator experiments
\cite{Porto,Matsumoto} and NMR experiments \cite{Sprague} measure
directly such basic quantities as the superfluid density, the pairing amplitude
and the spin susceptibility. 

In this letter we give theoretical explanations for some of the
experimental observations on superfluid $^3$He in aerogel. As a first attempt we
study a  model, where the aerogel is assumed as a
homogeneous scatterer of the quasiparticles of $^3$He. This model gives
predictions with a correct tendency, but it is insufficient quantitatively. A
``slab model'' gives a clue that the inhomogeneity of the scattering is
crucial for understanding the discrepancy. Based on that we construct a
relatively simple model of inhomogeneous scattering that quantitatively can
explain both the transition temperature and the pairing amplitude, and predicts
an inhomogeneity length scale of 100 nm.
We also consider the upper limit for anisotropic scattering set
by the NMR measurements.  

In the experiments the aerogel fills only 2\% of the total volume  ($V=0.02$),
and its surface to volume ratio is $A=260,000\ {\rm  cm}^{-1}$ \cite{KMC}.
Assuming naively that the material consists of a network of
one-dimensional strands, we can from these numbers alone estimate the strand 
diameter ${4V/ A}=3$ nm. The distance between strands is ${\sqrt{4\pi V}/A}=20$
nm. The mean free path for straight line trajectories is
estimated as $\ell={4/  A}=150$ nm. This is also the mean free path for
quasiparticles of $^3$He when the aerogel is filled with $^3$He at millikelvin
temperatures.

{\it Quasiclassical theory.}---Because the volume fraction of the aerogel
strands (including an inert layer of $^3$He atoms on the strands) is small, we
neglect all effects that are linear in the volume fraction. In particular, we
assume that the density, the Landau Fermi-liquid parameters, the coupling
constant of the pairing interaction, and the dipole-dipole interaction constant
are unchanged from the bulk. The changes of these parameters are of the same
order of magnitude as the volume fraction because they arise from processes of
relatively high energy and short length scale \cite{SRrev}. Much larger effects
on superfluidity arise from processes in the immediate vicinity of the Fermi
surface. Scattering of quasiparticles from the aerogel strands modifies the
superfluid state within the distance $\xi_0$, and causes an effect that is
proportional to the ratio $\xi_0/\ell$, which approaches unity in 98\%-porous
aerogel. Here $\xi_0$ is the superfluid coherence length. It is defined by
$\xi_0={\hbar v_{\rm F}/ 2\pi k_{\rm B}T_{\rm c0}}$, where $T_{\rm c0}$ is the 
transition temperature in bulk $^3$He and $v_{\rm F}$ the Fermi velocity. 
$\xi_0$ is a function of pressure varying between 16 nm (melting  pressure) and
77 nm (zero pressure).

Because of the $s$-wave pairing of conventional superconductors, their $T_{\rm
c}$ and paring amplitude is nearly unaffected by non-magnetic scattering at
$\xi_0/\ell\sim 1$, only the Ginzburg-Landau coherence length gets shorter
\cite{general}. But in a $p$-wave superfluid like $^3$He, the scattering causes
destructive interference and leads to complete depression of superfluidity
already at $\xi_0/\ell\sim 1$.

All the models we discuss are quasiclassical. This means that the aerogel is
modeled as a collection of incoherent scattering centers at locations ${\bf
r}_j$. Each center is assumed much smaller than $\xi_0$ but, similar to aerogel
strands, they can be large in comparison to the Fermi wave length $\lambda_{\rm
F}= 2\pi/k_{\rm F}\approx 0.7$ nm. For each scattering center, a fully
quantum-mechanical treatment is allowed in principle, but we describe them
phenomenologically by phase shifts and scattering cross sections.  The
interference of different scattering centers leads to weak localization
corrections, which are small because aerogel has random structure and 
$\lambda_{\rm F}/\ell\ll 1$. 

\begin{figure}[tbp]
\begin{center}\leavevmode
\includegraphics[width=0.80\linewidth]{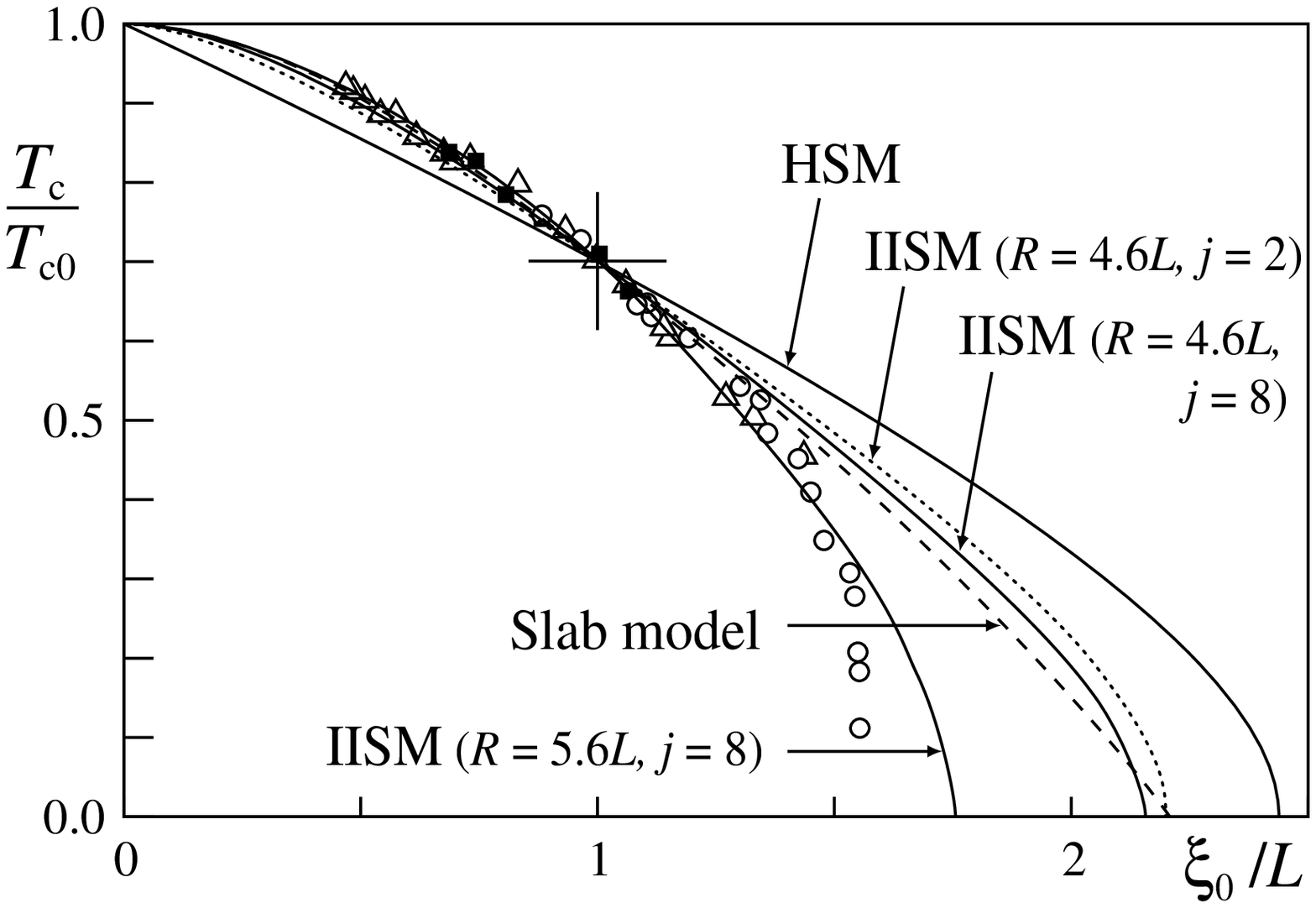}
\bigskip
\caption[f.tc]{ 
The transition temperature in aerogel relative to that in bulk,
$T_{\rm c}/T_{\rm c0}$. The horizontal axis is the coherence length
$\xi_0={\hbar v_{\rm F}/ 2\pi k_{\rm B}T_{\rm c0}}$ divided by $L$. The scale
$L$ is chosen so that the data sets coincide at the 
cross. The experimental results are from Refs.\ 
\cite{Porto} (\font\apu=cmsy7  {\apu\char'064}, $L=36$ nm),
\cite{Sprague} (\kern .2ex\vrule height .95ex width .8ex depth -.15ex
\kern .2ex, $L=25$ nm), and \cite{Matsumoto} ($\circ$, $L=24$
nm). The lines correspond to the homogeneous scattering model (HSM)
\cite{AG,larkin}, to the slab model \cite{KKR}, and to the isotropic
inhomogeneous scattering model (IISM) with different sphere radii $R$ 
and scattering profile parameters $j$. 
 }\label{f.tc}\end{center}\end{figure}

The coherence length $\xi_0$ is the only pressure dependent length scale in
scattering models \cite{pnote}. This implies that the calculated $T_{\rm c}$ can
be compared with experiments using the scaling presented in Fig.\
\ref{f.tc}. The vertical axis is the suppression of the transition temperature
relative to the bulk, $T_{\rm c}/T_{\rm c0}$. The horizontal axis is $\xi_0$
divided by a length $L$. The scale $L$ is a constant that characterizes
each aerogel sample. By definition, $L$ equals $\xi_0(p)$ at the pressure $p$
where $T_{\rm c}/T_{\rm c0}=0.7$. In other words, the horizontal scale is
chosen so that all data sets coincide at the point $(1.0,0.7)$. For the three
different samples used in the experiments we find $L=36$ nm
\cite{Porto}, $L=25$ nm \cite{Sprague}, and $L=24$ nm \cite{Matsumoto}. With
this scaling the three data sets seem rather consistent with each
other.

In order to compare the amplitude $\Delta(T,{\bf r})$ of the order parameter, we
study the suppression factor \cite{Sprague}
\begin{equation}
S_{\Delta^2}(t)={\langle\Delta^2(tT_{\rm c},{\bf r})\rangle\over
\Delta_0^2(tT_{\rm c0})}.
\label{e.sdelta}
\end{equation}
As before, the subscript $0$  refers to the bulk, {\it i.e.}\ to the case of
pure $^3$He. The parameter $t$ denotes the temperature relative to the
transition temperature. An average over locations ${\bf r}$  is indicated
by $\langle...\rangle$.

We can construct a suppression factor $S_{\rho_{\rm s}}$ for the superfluid
density
$\rho_{\rm s}$ in complete analogy with (\ref{e.sdelta}). However, $\rho_{\rm
s}$ depends strongly on the Fermi-liquid parameter $F_1^{\rm s}=3(m_{\rm
eff}/m-1)$. Because the pressure dependence of $F_1^{\rm s}$ spoils the scaling
with
$\xi_0$ discussed above, it is preferable to use the suppression factor
$S_{\tilde\rho_{\rm s}}$ for the bare superfluid density $\tilde\rho_{\rm s}$
defined by
$\rho_{\rm s}=\tilde\rho_{\rm s}/[1+{1\over 3}F_1^{\rm s}(1-\tilde\rho_{\rm
s}/\rho)]$, where
$\rho$ is the density of the liquid.

\begin{figure}[btp]
\begin{center}\leavevmode
\includegraphics[width=1.00\linewidth]{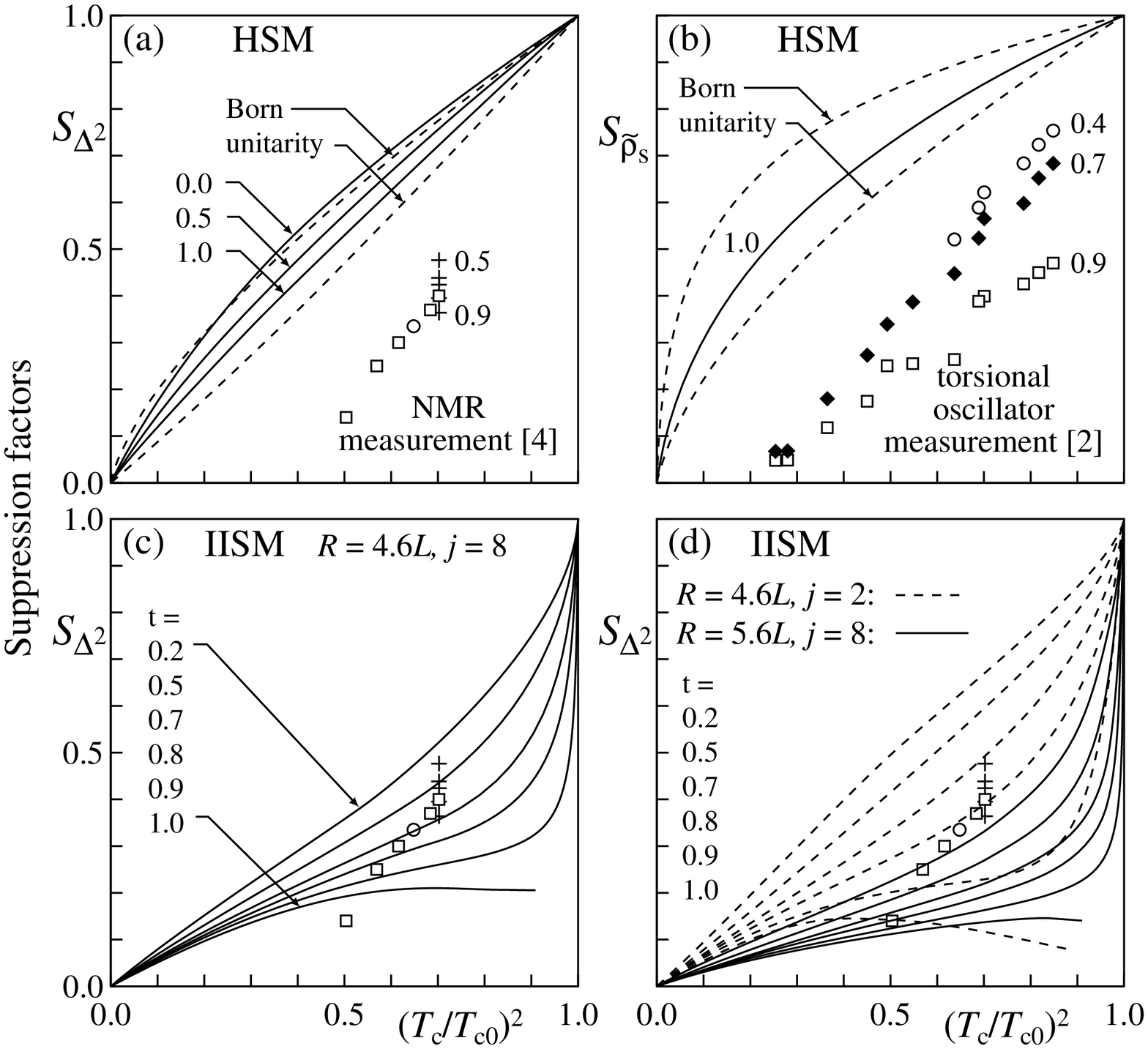}
\bigskip
\caption[f.s]{ 
The suppression factors for the gap ($S_{\Delta^2}$) and superfluid density
($S_{\tilde\rho_{\rm s}}$) as a function of squared $T_{\rm c}$ suppression,
$(T_{\rm c}/T_{\rm c0})^2$. The upper two frames present the experimental data
and the results of the homogeneous scattering model. The lower frames are
results of the isotropic inhomogeneous scattering model. The numbers
associated with curves and data points denote $t$, see equation
(\ref{e.sdelta}). Born ($\sin^2\delta_0\rightarrow 0$) and unitarity
($\sin^2\delta_0=1$) limits are shown by dashed lines in (a) and (b) at $t=1$.
All other curves are for the intermediate case
$\sin^2\delta_0=0.5$. }\label{f.s}\end{center}\end{figure}

The experimental suppression factors are plotted against $(T_{\rm c}/T_{\rm
c0})^2$ in Fig.\ \ref{f.s}(a, b). The NMR experiment measures $S_{\Delta^2}$
because the dipole-dipole interaction constant $g_{\rm d}$ \cite{Leggett} is
unchanged by scattering.
$S_{\tilde\rho_{\rm s}}$ can be extracted from torsional oscillator experiments.
The $t$ dependencies of $S_{\Delta^2}$ and $S_{\tilde\rho_{\rm s}}$ are
qualitatively similar, but it is more pronounced in
the latter.

{\it Homogeneous scattering model (HSM).}---This is the simplest scattering
model. The principal assumption is that the scattering probability
is independent of the location. Additionally we assume that the scattering
medium is isotropic, {\it i.e.}\
$\ell$ is independent of the direction of quasiparticle momentum. These are
just the standard assumptions made in studying impurities in superconductors
\cite{general}. We also neglect magnetic scattering because it does not seem to
be important for the effects we consider. A convenient property of the
isotropic HSM is that both the Ginzburg-Landau (GL) theory and Leggett's theory
of NMR \cite{Leggett} have the same form as in pure $^3$He. Only the parameters
of these theories have different values, as will be discussed below.

The Ginzburg-Landau theory is formulated in terms of a free  
energy 
functional of the $3\times 3$ matrix order parameter, $A_{\mu 
i}$, where $\mu$ represents the spin components and $i$  
represents
the orbital components of the pair state. The ``bulk'' terms  are
\cite{T87,3Hereview} 
\begin{eqnarray}
f_{\rm bulk}&=&\alpha A_{\mu i}^*A_{\mu i}
+\beta_1\vert A_{\mu i}A_{\mu i}\vert^2
+\beta_2(A_{\mu i}A_{\mu i}^*)^2\nonumber \\\mbox{}&&
+\beta_3 A_{\mu i}^*A_{\nu i}^*A_{\nu j} A_{\mu j}
+\beta_4 A_{\mu i}^*A_{\nu i}A_{\nu j}^*A_{\mu j}
\nonumber \\\mbox{}&&
+\beta_5 A_{\mu i}^*A_{\nu i}A_{\nu j}A_{\mu j}^*\ .
\label{e.gl}
\end{eqnarray}
The transition temperature $T_{\rm c}$ is determined by the condition 
$\alpha(T_{\rm c})=0$. Minimizing (\ref{e.gl}) one finds the order 
parameter amplitudes $\Delta$ and free energies $f$ of the various 
phases. For example, the polar, planar, and B
phases have $f=k\alpha\Delta^2/2=- 
k\alpha^2/(4k\beta_{12}+4\beta_{345})$ with $k=1$, 2,
and 3, respectively, where $\beta_{ij...}=\beta_i+\beta_j+...$, and the 
A phase has $f=\alpha\Delta^2=- \alpha^2/4\beta_{245}$.

The coefficient $\alpha$ is given by \cite{larkin}
\begin{equation}
\alpha={N(0)\over 3}\left[\ln{T\over 
T_{\rm c0}}+\sum_{n=1}^\infty\left({1\over n-{1\over 2}}-{1\over
n-{1\over 2}+x}\right)\right],\label{e.alpha}
\end{equation}
where $x=\hbar v_F/4\pi T\ell_{\rm tr}$, $\ell_{\rm tr}$ is the transport 
mean free path, and $2N(0)$ the density of states at the Fermi surface. 
The suppression of $T_{\rm c}$ in the HSM is shown in Fig.\ \ref{f.tc}
($\ell_{\rm tr}=8.7L$). Its dependence on $\xi_0$ is the same as found for
magnetic impurities in
$s$-wave superconductors \cite{AG}.

For the  coefficients $\beta_i$ we make the additional
assumption that only $s$-wave scattering is important, and obtain
\begin{eqnarray}
\left(\matrix{\beta_1\cr \beta_2\cr \beta_3\cr \beta_4\cr
\beta_5\cr}\right)&=&a\left(\matrix{-1/2\cr 1\cr 1\cr 1\cr
-1\cr}\right)+b\left(\matrix{0\cr 1\cr 0\cr 1\cr -1\cr}\right)
+\left(\matrix{\Delta\beta_1^{\rm sc}\cr \Delta\beta_2^{\rm sc}\cr 
\Delta\beta_3^{\rm sc}\cr \Delta\beta_4^{\rm sc}\cr
\Delta\beta_5^{\rm sc}\cr}\right)
\label{e.beta}\\
a&=&{N(0)\over 120(\pi T)^2}\sum_{n=1}^\infty(n-{\textstyle{1\over 2}}+x)^{-
3}\nonumber \\
b&=&{N(0)v_{\rm F}\over 288(\pi T)^3\ell }\left(\sin^2\delta_0-
{\textstyle{1\over 2}}\right)
\sum_{n=1}^\infty(n-{\textstyle{1\over 2}}+x)^{-4}.\nonumber 
\end{eqnarray}
Besides $\ell$, $b$ also depends on the $s$-wave scattering phase
shift $\delta_0$. The effect of this fully quantum-mechanical degree of
freedom on the suppression factors is shown by dashed lines in Fig.\
\ref{f.s}(a, b). However, calculations taking into account higher partial waves
show that this dependence is essentially averaged out for large scatterers
\cite{TKR}. The end result is approximately the same as if the phase shifts were
random: $\sin^2\delta_0\rightarrow 0.5$. Therefore we
chose $\sin^2\delta_0= 0.5$ in all other results of the HSM and IISM.

The suppression factors of the HSM are essentially the same for different
superfluid phases. The difference between A and B phases is negligible also in
$S_{\tilde\rho_{\rm s}}$, when the average of the tensor $\tensor\rho_{\rm s}$
is used for the A phase.  The stability of A and B phases depends on strong
coupling corrections
$\Delta\beta_i^{\rm sc}$, which are not known. Assuming they remain constants,
the B phase is favored by increasing scattering. No new phases are stabilized.

We conclude the HSM by noting that it works in the right direction for  
all $T_{\rm c}$, $\Delta$, and $\rho_{\rm s}$, but quantitatively, on the level
of accuracy we are accustomed to in superfluid $^3$He, it is clearly
inadequate. 

{\it Slab model.}---This model considers $^3$He in a gap of thickness $D$
between two diffusely scattering planes. The dashed line in Fig.\ \ref{f.tc}
shows $T_{\rm c}$ calculated in Ref.\ \cite{KKR} ($D=2.95L$). The agreement with
measurement is much better than for the HSM. In particular, $T_{\rm c}$
suppression is quadratic at small $\xi_0$ compared to linear in the HSM.
Generally, this feature arises from regions that have no scattering nearby,
such as the center of the slab. The suppression of $\langle\Delta^2\rangle$
also is in better agreement with experiments than in the HSM \cite{LT21}. The
principal deficiency of the slab model is its strong anisotropy. We estimate
(see below) that in order to be in agreement with the measured NMR shift, the
normal direction of the slab has to vary randomly on a length scale that is
smaller than the thickness
$D$. This contradiction prompts us to look for a better model.  

{\it Isotropic inhomogeneous scattering model (IISM).}---Experimentally
the suppression of $T_{\rm c}$ seems quadratic at small $\xi_0$ (Fig.\
\ref{f.tc}). This implies that real aerogel has voids, {\it i.e.}\ regions of
negligible scattering. In order to construct a model that is feasible in
calculations, we make two basic simplifications. (i) Instead of a
random distribution of voids, we consider a periodic lattice of them. (ii) The
unit cell of this lattice is approximated by a sphere. In more detail,
the boundary condition is that a quasiparticle escaping from the
sphere will be returned there at the diametrically opposite point but its
momentum is unchanged. (A phase shift similar as in Bloch wave functions should
be added in the case of nonconstant phase.) In addition to the radius $R$ of the
sphere, we need to specify how the density
$n({\bf r})$ of scattering centers is distributed in the sphere. When $n({\bf
r})$ depends only on the radial coordinate, {\it i.e.}\  $n({\bf r})=n(r)$, the
model is completely isotropic. We study polynomial forms
$n(r)=c[(r/R)^j-j(r/R)^{j+2}/(j+2)]$. We show results in Figs.\ \ref{f.tc}
and \ref{f.s}(c, d) for a steep ($j=8$) and a
slow ($j=2$) profile. The suppression factors are for an inhomogeneously
distorted B phase, but extrapolating the experience from two previous models,
the A-type phase would be quite similar.

In spite of the inhomogeneity, there is a single well defined $T_{\rm c}$ at
which the order parameter becomes nonzero. Due to the proximity effect, $T_{\rm
c}$ is determined collectively by the whole sample, although the weight
of high-scattering regions far from low-scattering ones is exponentially
small. Anyway, the transition can be described as ``broadened'' if
$\langle\Delta^2\rangle(T)$ is strongly nonlinear below $T_{\rm c}$. This is
the case for a slow profile [dashed lines in Fig.\ \ref{f.s}(d)] and
for large $T_{\rm c}/T_{\rm c0}$: in Fig.\ \ref{f.s}(c, d) this shows up as
strong $t$ dependence of $S_{\Delta^2}$. In contrast,
$\langle\Delta^2\rangle(T)$ is nearly linear over the whole temperature range
for small $T_{\rm c}/T_{\rm c0}$. [In this case, the $t$ dependence of
$S_{\Delta^2}$ arises mostly from nonlinearity of the reference
$\Delta_0^2(T)$.]

The IISM reduces to the HSM in the limit of small $R$. This means that the
true distribution of the scattering centers is irrelevant as long
as the average scattering over a length scale $\xi_0$ remains the same
\cite{FKBS}. 

We see that the IISM is in much better agreement with experiments than
the HSM when $R\approx 5L$ and $j\approx 8$. The magnitude and $t$
dependence of $S_{\Delta^2}$ and most of $T_{\rm c}(\xi_0)$ is well accounted
for. There is a small systematic deviation 
that experimentally both $T_{\rm c}$ and $S_{\Delta^2}$ drop more rapidly
with increasing $\xi_0$ than in the model. We believe that this difference
arises from the periodicity assumption in the IISM. In real aerogel there are
fluctuations of all length scales, and with increasing
$\xi_0$ the length scale of most relevant fluctuations also increases. This
is consistent with the observed deviations which imply an increasing
effective $R$ for increasing $\xi_0$. 

The $\rho_{\rm s}(T)$ measurement [Fig.\ \ref{f.s}(b)] shows considerably more
nonlinearity than the NMR measurement [Fig.\ \ref{f.s}(a)].
A possible explanation for this is that the structure of the aerogel is
different in the two experiments, the former corresponding to a smaller $j$. In
order to confirm such a hypothesis, both samples should be studied with the
same measuring technique.

The large scattering fluctuations predicted by the IISM seem to us 
natural and essentially unique explanation of the measured suppression factors.
However, the predicted effective void radius $\approx 0.8R\sim 100...150$ nm is
very large compared to the estimates based on small angle x-ray scattering
\cite{Porto}. This problem remains open. 

{\it Anisotropic HSM.}---According to our introductory estimate, the aerogel
consists of randomly oriented strands of diameter 3 nm and length $L_{\rm a}
\approx 20$ nm. This kind of anisotropy can have several consequences on the
superfluid state. It is known that strongly anisotropic scattering, such as in
the slab model, can stabilize the A phase, and one can ask if the aerogel
strands could do the same. We have studied this in the limit
$L_{\rm a}\lesssim\xi_0$, where we recover the GL functional (\ref{e.gl}) of the
isotropic HSM since the anisotropy is averaged out on the scale $\xi_0$.
However, the coefficients $\beta_i$ are modified. We find that anisotropic
backscattering, preferentially perpendicular to the strands, can stabilize the
A-phase low pressures, where the B-phase is otherwise stable\cite{HSM+}.

The anisotropy couples to the orbital part of the order parameter. For example,
the $\alpha$-term in the GL functional (\ref{e.gl}) is modified to $\alpha_{ik}
A_{\mu i}^*A_{\mu k}$. Let us study this in the A phase, where the orbital and
spin parts are described by $\hat{\bf l}$ and $\hat{\bf d}$ vectors,
respectively. There are two possibilities \cite{Volovik}. (i) For weak
anisotropy the dipole-dipole coupling between $\hat{\bf l}$ and $\hat{\bf d}$
keeps them aligned to each other. (ii) For strong anisotropy, $\hat{\bf l}$ is
driven to vary randomly on a scale $L_{\rm o}$ where $\hat{\bf d}$ is still
nearly constant. Using similar estimates than Imry and Ma \cite{IM} we find
that NMR frequency shift in the former state is unchanged relative to HSM, but
it is reduced to essentially zero in the latter. Experiments clearly point to
the former state \cite{Sprague}. Our scattering estimate also favors this state,
but the margin is rather small: if $L_{\rm a}\approx 50$ instead of $20$ nm, the
latter state would be favored. The proximity of the transition gives natural
explanation to the observed sudden extinction of the NMR shift as a function
tipping angle
\cite{Sprague,Volovik}. 

{\it Conclusions.}---Superfluid $^3$He in aerogel is in many respects an ideal
system to study impurity effects in unconventional superfluidity. We find,
in particular, that the standard impurity model is robust in the sense that
large fluctuations in the scattering are needed in order to get such
substantial deviations as seen experimentally. 

We thank R. H\"anninen and T. Set\"al\"a for help in numerical calculations.


\begin{thebibliography}{10}
\bibitem{pnote} Here we neglect all strong-coupling corrections \cite{SRrev}.
The direct dependence of the cross section $\sigma$ of a scattering center
on the Fermi momentum is estimated small for objects which are large in
comparison to
$\lambda_{\rm F}$. There remains, though, an uncertainty how the wave function
of a quasiparticle, which is not known, changes with pressure and affects
$\sigma$.
\end{thebibliography}
\end{document}